\documentclass[a4paper,11pt]{article}
\pdfoutput=1 %

\usepackage{jheppub} 
\usepackage{amsmath,amssymb,graphicx,float,slashed,xcolor,multicol}
\usepackage{tabularx}
\usepackage{url}
\usepackage{footmisc}
\usepackage{amsfonts}
\usepackage{cancel}
\usepackage{color}
\usepackage{multirow} 
\usepackage{pifont}
\usepackage{epstopdf}
\usepackage{comment}
\usepackage{booktabs} 
\usepackage{natbib}
\usepackage{array}
\usepackage{mathrsfs}
\usepackage[toc,page]{appendix}
\usepackage{mathtools}
\usepackage{romannum}
\usepackage{ulem}
\usepackage{bbold}
\usepackage{enumitem}
\usepackage{multirow}
\usepackage{cleveref}
\usepackage{caption,subcaption}
\usepackage{float}
\usepackage{multirow}
\usepackage{upgreek}
\usepackage[toc,page]{appendix}
\usepackage{romannum}
\allowdisplaybreaks
\usepackage{bbm}                
\usepackage{xspace}				

\usepackage{tikz}
\usetikzlibrary{arrows,shapes}
\usetikzlibrary{trees}
\usetikzlibrary{matrix} 
\usetikzlibrary{positioning}				
\usetikzlibrary{calc,through}				
\usetikzlibrary{decorations.pathreplacing}  
\usepackage{pgffor}							
\usetikzlibrary{decorations.pathmorphing}	
\usetikzlibrary{decorations.markings}
\tikzset{
	vector/.style={decorate, decoration={snake}, draw},
	provector/.style={decorate, decoration={snake,amplitude=2.5pt}, draw},
	antivector/.style={decorate, decoration={snake,amplitude=-2.5pt}, draw},
	fermion/.style={draw=black, postaction={decorate},
		decoration={markings,mark=at position .55 with {\arrow[draw=black]{>}}}},
	fermionbar/.style={draw=black, postaction={decorate},
		decoration={markings,mark=at position .55 with {\arrow[draw=black]{<}}}},
	fermionnoarrow/.style={draw=black},
	gluon/.style={decorate, draw=black,
		decoration={coil,amplitude=4pt, segment length=5pt}},
	scalar/.style={dashed,draw=black, postaction={decorate},
		decoration={markings,mark=at position .55 with {\arrow[draw=black]{>}}}},
	scalarbar/.style={dashed,draw=black, postaction={decorate},
		decoration={markings,mark=at position .55 with {\arrow[draw=black]{<}}}},
	scalarnoarrow/.style={dashed,draw=black},
	electron/.style={draw=black, postaction={decorate},
		decoration={markings,mark=at position .55 with {\arrow[draw=black]{>}}}},
	bigvector/.style={decorate, decoration={snake,amplitude=4pt}, draw},
}

\tikzset{>=latex} 

\usepackage{tikz-feynman}
\makeatletter
\tikzfeynmanset{compat=\tikzfeynman@version@major.\tikzfeynman@version@minor.\tikzfeynman@version@patch}
\makeatother

\tikzstyle{block} = [draw, rectangle, 
minimum height=3em, minimum width=6em]

\newcommand*{\rom}[1]{\expandafter\@slowromancap\romannumeral #1@}

\newcommand{\Q}{\mathcal {O}}
\newcommand{\C}{\mathcal {C}}

\def\lag{\mathscr{L}}

\def\us{\underset}

\def\beq{\begin{equation}}
\def\eeq{\end{equation}}
\def\beqa{\begin{eqnarray}}
\def\eeqa{\end{eqnarray}}
\title{Anomalous dimensions from Yukawa couplings in SMNEFT: four-fermion operators}
\author[a]{Alakabha Datta,}
\author[b]{Jacky Kumar,}
\author[c]{Hongkai Liu,}
\author[d]{Danny Marfatia}
\affiliation[a]{Department of Physics and Astronomy, University of Mississippi, Oxford, MS 38677, USA}
\affiliation[b]{Institute for Advanced Study, Technical University of Munich,
85748 Garching, Germany}
\affiliation[c]{Department of Physics and Astronomy, University of Pittsburgh, Pittsburgh, PA 15260, USA}
\affiliation[d]{Department of Physics and Astronomy, University of Hawaii at Manoa, Honolulu, HI 96822, USA}
\emailAdd{datta@phy.olemiss.edu}
\emailAdd{jacky.kumar@tum.de}
\emailAdd{hol42@pitt.edu}
\emailAdd{dmarf8@hawaii.edu}

\preprint{
\begin{flushright}
\end{flushright}
}

\abstract{ The Standard Model Neutrino Effective Field Theory (SMNEFT) is the Standard Model Effective Field Theory (SMEFT)  augmented with
	right-handed neutrinos. Building on our previous work, arXiv:2010.12109, we calculate the Yukawa coupling contributions to the one-loop anomalous dimension matrix for the 11 dimension-six four-fermion SMNEFT operators. We also present the new contributions to the anomalous dimension matrix for the 14 four-fermion SMEFT operators that mix with the SMNEFT operators through the Yukawa couplings of the right-handed neutrinos.
}

\begin{document}

\titlepage
\maketitle


\flushbottom

\section{Introduction}
\label{sec:intro}

The sterile right-handed neutrino is one of the most well studied extensions of the Standard Model (SM) motivated by, among other things, the observation of neutrino masses and mixing. 
Instead of considering all possible models, an efficient alternative is to  use a model independent approach based on the principles of effective theory. The idea is to construct all possible operators representing the interactions of sterile neutrinos with SM fields consistent with the symmetries of the SM. The framework  is valid
between  the scale of electroweak symmetry breaking, $\mu_{EW}$, and  the cut-off scale for new physics, $\Lambda$.

In the effective theory approach, the leading 
terms of the effective Lagrangian are given by the SM, and  the new interactions of the right-handed neutrino with the SM fields
 are described by higher dimension
operators,
\begin{eqnarray}
{\cal L}  =  \sum_{i} {\cal{C}}_i {\cal O}_i\,. \ 
\end{eqnarray}
The operators ${\cal O}_i$ respect the
	$SU(3)_C \times SU(2)_L \times U(1)_Y$ gauge symmetry and are constructed from SM and right-handed neutrino fields. The renormalization scale dependent Wilson coefficient (WC)
 ${\cal {C}}_i$,
determines the size of the contribution of operator ${\cal O}_i$,
and is calculated by matching the effective theory with the underlying theory.

Given the absence of new physics signals at the
	LHC, the use of effective theory to study physics beyond the  SM has received much attention recently. With only SM fields,
the Standard Model Effective Field Theory (SMEFT) is obtained~\cite{Buchmuller:1985jz, Grzadkowski:2010es,Henning:2014wua, Brivio:2017vri},
and the one-loop renormalization group evolution (RGE) of all dimension-six operators have been presented in Refs.~\cite{Jenkins:2013zja, Jenkins:2013wua, Alonso:2013hga}. 

Extending SMEFT with sterile right-handed neutrinos $n$, yields the Standard Model Neutrino Effective Field Theory (SMNEFT)~\cite{delAguila:2008ir, Aparici:2009fh, Bhattacharya:2015vja, Liao:2016qyd, Bischer:2019ttk}.  Loop effects in SMNEFT have only recently started being studied. 
We presented the gauge terms of the one-loop RGE of all dimension-six operators in SMNEFT~\cite{Datta:2020ocb}.
  The mixing between the bosonic operators was discussed in Ref.~\cite{Chala:2020pbn}, and the one-loop RGE of a subset of four-fermion operators was provided in Ref.~\cite{Han:2020pff}. 
 
In this paper we calculate the one-loop RGE of all dimension-six four-fermion SMNEFT operators that arises from the Yukawa interactions of the Higgs, right-handed neutrino and SM fields. We assume the neutrinos are Dirac in nature. 
Our formalism is generalizable to the Type-I two-Higgs-doublet model.  Note that the neutrino Yukawa couplings may be large if a contribution to Dirac neutrino masses from high scale physics, as for example in Ref.~\cite{Cvetic:2008hi},  is rendered small by virtue of a cancellation by the Yukawa contribution.
We present our calculations and results in the gauge basis
because a transformation
to the mass basis involves a rotation of the quark and lepton fields to their mass eigenstates. This rotation is inherently model-dependent since only the left-handed quark and lepton mixing matrices are experimentally accessible. 
Moreover, how the Yukawa interactions of the right-handed neutrinos relate to the neutrino mixing parameters depends on the mechanism of neutrino mass generation. 

The paper is organized as follows. In section~2, we introduce the formalism to compute the RGE of the dimension-six operators. In section~3, we present the  one-loop anomalous dimension matrix (ADM) of all four-fermion operators in SMNEFT, and the additional RGE terms in SMEFT that arise from Yukawa couplings of the right-handed neutrinos. Finally, in section~4 we present our summary.


\section{Formalism}

\begin{figure}[t]
	\centering
	\begin{subfigure}{.24\textwidth}
		\centering
		\includegraphics[width=\textwidth]{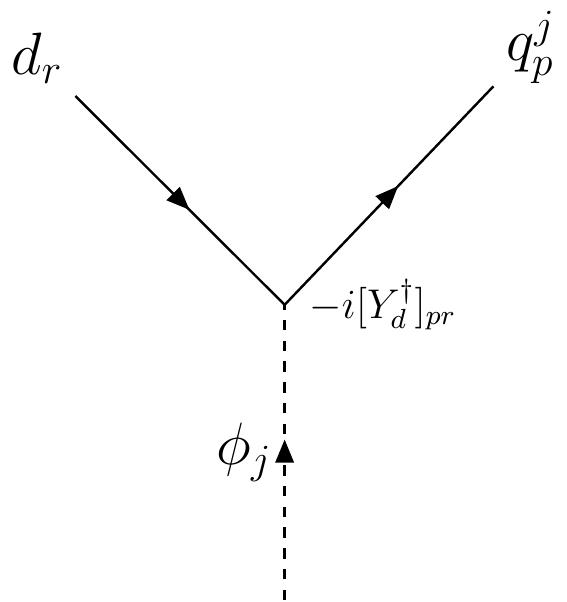}
	\end{subfigure}
	\begin{subfigure}{.24\textwidth}
		\centering
		\includegraphics[width=\textwidth]{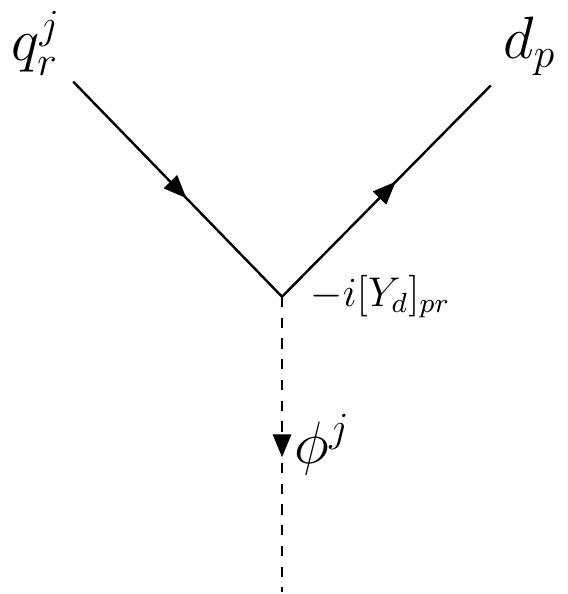}
	\end{subfigure}
	\begin{subfigure}{.24\textwidth}
		\centering
		\includegraphics[width=\textwidth]{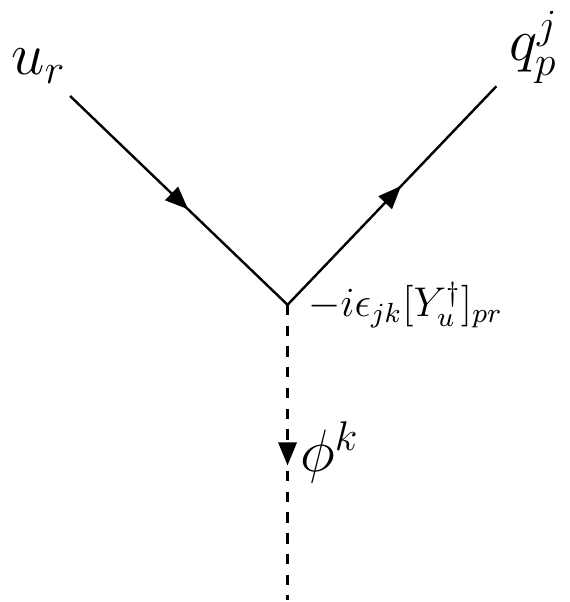}
	\end{subfigure}
	\begin{subfigure}{.24\textwidth}
	\centering
	\includegraphics[width=\textwidth]{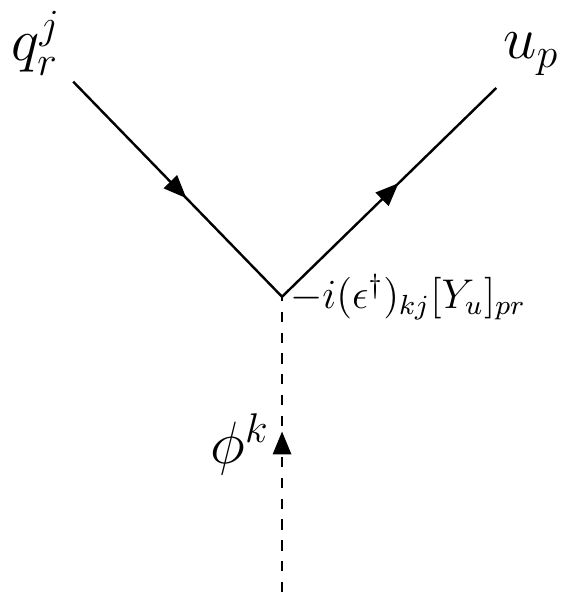}
\end{subfigure}
	\caption{The four types of Yukawa interaction vertices. The flavor indices `$pr$' and $SU(2)_L$ indices `$jk$' are written explicitly. 
	}
	\label{fig:Yukawa_vertex}
\end{figure}

The SMNEFT Lagrangian is
\beq
\lag_{\text{SMNEFT}} \supset  i\bar{n}\slashed{\partial}n+\lag_{\rm{Yukawa}}+ \sum_{i}\C_i \Q_i\,,
\eeq
where $\C_i$ are the WCs of the dimension-six operators. The Yukawa terms with generation indices suppressed, are
\beq
\lag_{\text{Yukawa}} = - [\phi^{\dagger j}\bar{d} Y_d q_j + \tilde{\phi}^{\dagger j}\bar{u} Y_u q_j + \phi^{\dagger j}\bar{e} Y_e \ell_j + \tilde{\phi}^{\dagger j}\bar{n} Y_n \ell_j + \text{h.c.}]~,
\eeq
where $\phi$ is the Higgs doublet and $\tilde{\phi}^j = \epsilon^{jk} \phi_k^*$. The four types of Yukawa interaction vertices for the quark sector are shown in Fig.~\ref{fig:Yukawa_vertex}.
There are a total of 16 ($\Delta B = 0 = \Delta L $) new operators in the SMNEFT framework, which are shown in Table~\ref{Tb:SMNEFT} in the Warsaw basis convention~\cite{Grzadkowski:2010es}.

The Lagrangian can be written in terms of bare fields $\vec{\Q}^{(0)}$ as
\beq
\lag_{\rm{SMNEFT}}\supset \vec{\C}^{~T} \cdot \vec{\Q}  +{\textit{\rm counterterms}} 
\equiv \vec{\C}^{~T}\cdot \mathbf{Z}\cdot \vec{\Q}^{(0)}\,,
\eeq
where $\mathbf{Z} = \mathbf{Z}_{\text{ct}}/\mathbf{Z}_{\text{wr}}$ is the renormalization constant matrix which depends on corrections from the counterterms,
 $\mathbf{Z}_{\text{ct}}$, and the wavefunction renormalizations, $\mathbf{Z}_{\text{wr}}$. Given that the bare operators and Lagrangian are independent of the renormalization scale $\mu$, the RG equations for the Wilson coefficients are
\beq
\dot{\vec{\C}}\equiv 16\pi^2\mu \frac{d}{d\mu}\vec{\C}=  - 16\pi^2 (\mathbf{Z}^T)^{-1} \mu \frac{d}{d\mu} \mathbf{Z}^T \vec{\C}\,.
\eeq
The main task is to calculate the  expressions for $\mathbf{Z}_{\text{wr}}$ and $\mathbf{Z}_{\text{ct}}$, which is detailed below.

\begin{table}
	\centering
	\renewcommand{\arraystretch}{1.5}
	\begin{tabular}{|c|c|c|c|c|c|}
		\hline
		\multicolumn{2}{|c|}{$(\bar RR)(\bar RR)$} &
		\multicolumn{2}{|c|}{$(\bar LL)(\bar RR)$} &
		\multicolumn{2}{|c|}{$(\bar LR)(\bar RL)$ and $(\bar LR)(\bar LR)$ }\\
		\hline
		$\Q_{nd}$  & $(\bar n_p \gamma_\mu n_r)(\bar d_s \gamma^\mu d_t)$ &
		$\Q_{qn}$  & $(\bar q_p \gamma_\mu q_r)(\bar n_s \gamma^\mu n_t)$ &
		$\Q_{\ell n \ell e}$  & $(\bar \ell^j_p  n_r)\epsilon_{jk}(\bar \ell^k_s e_t)$   \\
		$\Q_{nu}$  & $(\bar n_p \gamma_\mu  n_r)(\bar u_s \gamma^\mu  u_t)$ &
		$\Q_{\ell n}$ & $(\bar \ell_p \gamma_\mu \ell_r)(\bar n_s\gamma^\mu n_t)$ &
		$\Q_{\ell n q d}^{(1)}$ &  $(\bar \ell^j_p  n_r)\epsilon_{jk}(\bar q^k_s  d_t)$  \\
		$\Q_{ne}$  & $(\bar n_p \gamma_\mu n_r)(\bar e_s \gamma^\mu e_t)$ &
		& &
		$\Q_{\ell n q d}^{(3)}$  & $(\bar \ell^j_p  \sigma_{\mu\nu} n_r)\epsilon_{jk}(\bar q^k_s \sigma^{\mu\nu} d_t)$     \\
		$\Q_{nn}$  & $(\bar n_p \gamma_\mu n_r)(\bar n_s \gamma^\mu  n_t)$ &
		&  &
		$\Q_{\ell n u q}$  & $(\bar \ell^j_p  n_r)(\bar u_s q^j_t)$  \\
		$\Q_{nedu}$      & $(\bar n_p \gamma_\mu e_r)(\bar d_s \gamma^\mu  u_t)$   & & &
		&   \\
		\hline
		\multicolumn{2}{|c|}{$\psi^2\phi^3$} &
		\multicolumn{2}{|c|}{$\psi^2\phi^2 D$} &
		\multicolumn{2}{|c|}{$\psi^2 X \phi$}\\
		\hline
		$\Q_{n\phi}$  & $(\phi^\dagger \phi) (\bar l_p  n_r \tilde \phi)$ & $\Q_{\phi n}$   &
		$i(\phi^\dagger \overset{\leftrightarrow}{D}_\mu \phi) (\bar n_p \gamma^\mu n_r) $& $\Q_{nW}$ & $(\bar \ell_p  \sigma^{\mu\nu} n_r)\tau^I \tilde \phi W_{\mu\nu}^I$ \\
		&  & $\Q_{\phi ne}$   &  $i(\tilde \phi^\dagger D_\mu \phi) (\bar n_p \gamma^\mu e_r) $  & $\Q_{nB}$  & $(\bar \ell_p  \sigma^{\mu\nu} n_r)\tilde \phi B_{\mu\nu}$ \\
		&  &  &  &  &  \\
		\hline
	\end{tabular}
	\bigskip
	\captionsetup{width=0.9\textwidth}
	\caption{\small \label{tab:SMNEFTops} The 16 SMNEFT operators involving the right-handed neutrinos $n$ in the 
		Warsaw convention which conserve baryon and lepton number $(\Delta B=\Delta L=0)$. 
		The flavor indices `$prst$' are suppressed for simplicity. The fundamental $SU(2)_L$ 
		indices are denoted by $j,k$, and $I$ is the adjoint index. 
		}
	\label{Tb:SMNEFT}
\end{table}

\subsection{Wavefunction renormalization}

\begin{figure}[t]
	\centering
		\includegraphics[width=.5\textwidth]{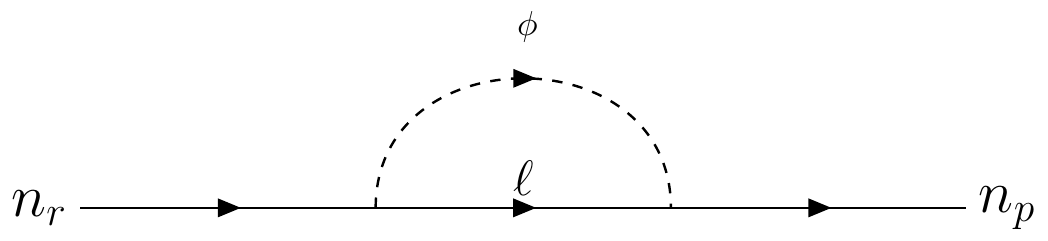}
		\caption{Self-energy of right-handed neutrino $n$. }
		\label{fig:SE_n}
\end{figure}

The bare field $\psi^{(0)}$ is related to the renormalized field $\psi^{R}$ via
\beq
\psi^{R} = \frac{1}{\sqrt{Z_\psi}}\psi^{(0)}.
\eeq
For the four-fermion operator, $\Q_{4\psi} = \bar\psi_1\psi_2\bar\psi_3\psi_4$, the wavefunction renormalization constant  is given by
\beq
\mathbf{Z}_{\text{wr}}\equiv\sqrt{\prod_{i=1}^{4} \mathbf{Z}_{\psi_i}}\,.
\eeq
From Fig.~\ref{fig:SE_n}, the Yukawa-dependent wavefunction renormalization of a right-handed neutrino $n$ is
\beq
Z^{(Y)}_{\us{pr}{n}} = 1- \frac{\gamma_{\us{pr}{n}}^{(Y)} }{16\pi^2\epsilon}\,,
\eeq
where $\gamma_{\us{pr}{n}}^{(Y)} = [Y_n Y^\dagger_n]_{pr}$, in the notation of Ref.~\cite{Jenkins:2013wua}.  We have taken dimension, $D = 4 - 2 \epsilon$, in dimensional regularization. Similarly,
we have,
\beqa
\gamma_{\us{pr}{\ell}}^{(Y)} &=& \frac{1}{2} [Y_e^\dagger Y_e + Y^\dagger_n Y_n]_{pr}\,,\quad\gamma_{\us{pr}{e}}^{(Y)} = [Y_e Y^\dagger_e]_{pr}\,,\nonumber\\
\gamma_{\us{pr}{q}}^{(Y)} &=& \frac{1}{2} [Y^\dagger_d Y_d + Y^\dagger_u Y_u]_{pr}\,,\quad\gamma_{\us{pr}{d}}^{(Y)} = [Y_d Y^\dagger_d]_{pr},\quad\gamma_{\us{pr}{u}}^{(Y)} = [Y_u Y^\dagger_u]_{pr}\,.
\eeqa

	\begin{figure}[t]
	\centering
	\begin{subfigure}{0.24\textwidth}
		\includegraphics[width=\textwidth]{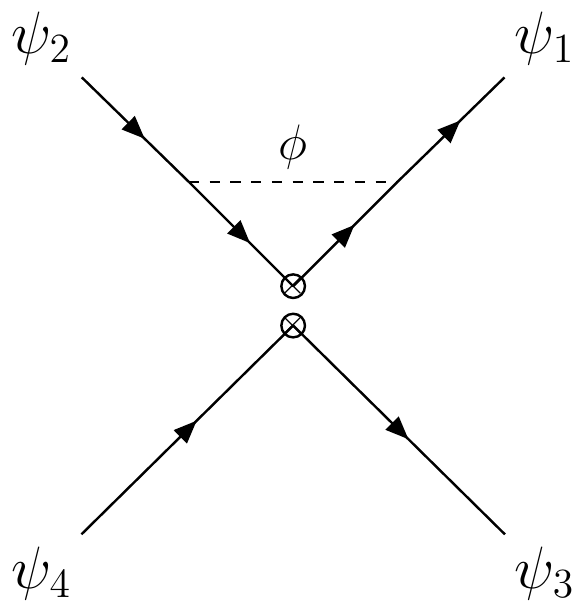}
		\caption{}
		\label{fig:4f_12}
	\end{subfigure}
	\begin{subfigure}{0.24\textwidth}
		\includegraphics[width=\textwidth]{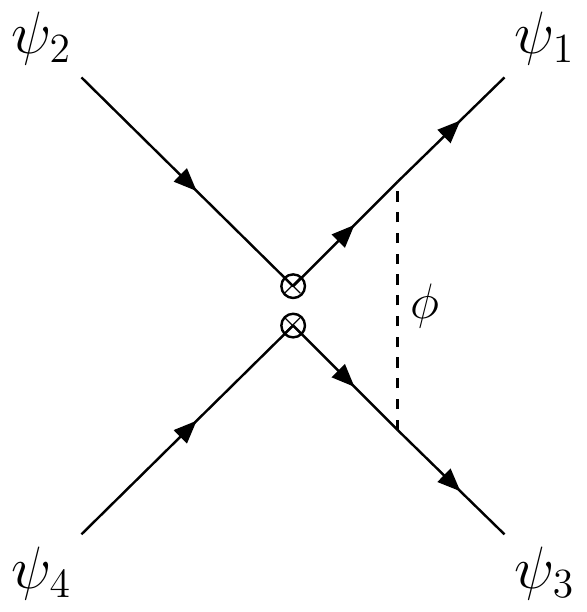}
		\caption{}
		\label{fig:4f_13}
	\end{subfigure}
	\begin{subfigure}{0.24\textwidth}
		\includegraphics[width=\textwidth]{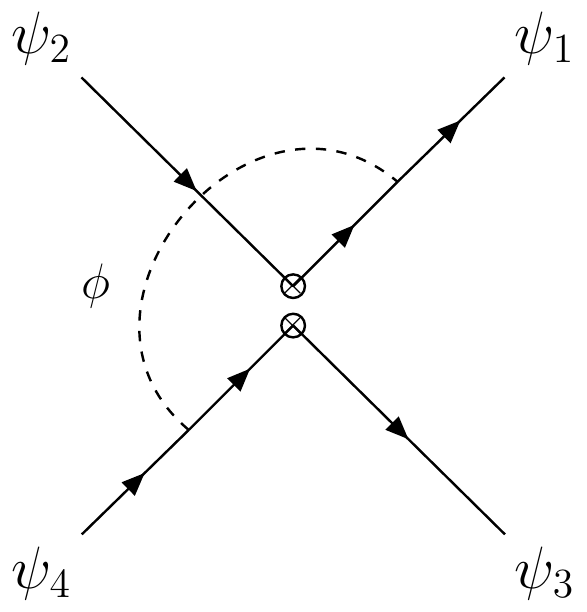}
		\caption{}
		\label{fig:4f_14}
	\end{subfigure}
	\begin{subfigure}{0.175\textwidth}
		\includegraphics[width=\textwidth]{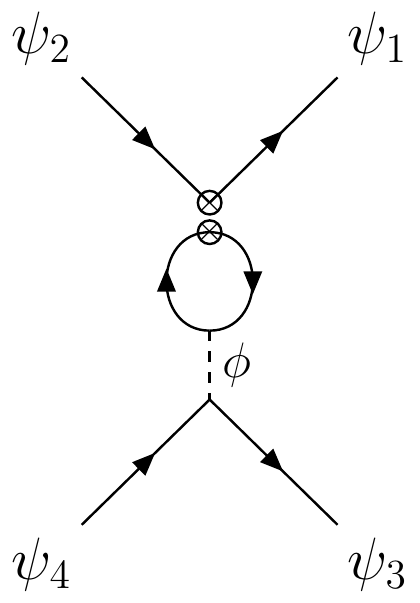}
		\caption{}
		\label{fig:4f_Xi}
	\end{subfigure}
	\begin{subfigure}{0.24\textwidth}
		\includegraphics[width=\textwidth]{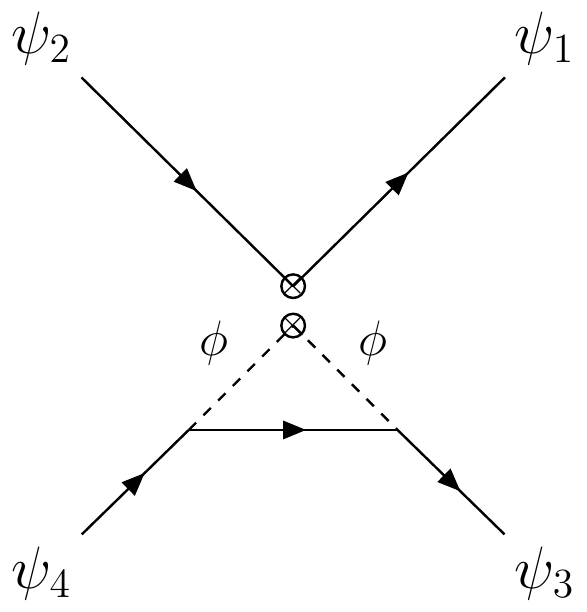}
		\caption{}
		\label{fig:4f_Hf}
	\end{subfigure}
	\begin{subfigure}{0.24\textwidth}
		\includegraphics[width=\textwidth]{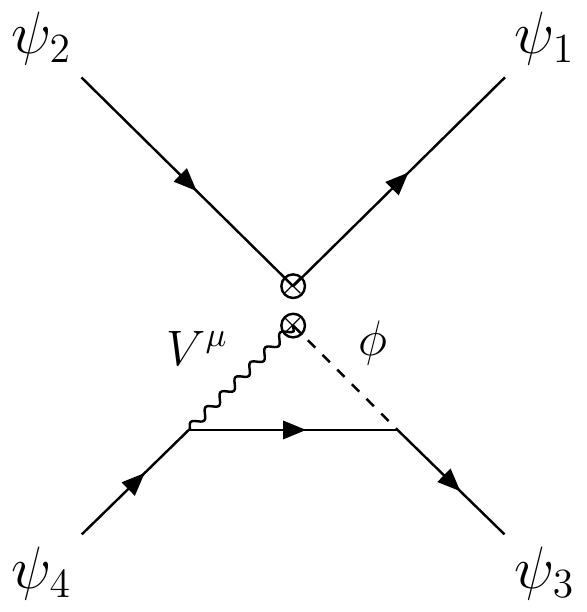}
		\caption{}
		\label{fig:4f_HA}
	\end{subfigure}
	\begin{subfigure}{0.24\textwidth}
		\includegraphics[width=\textwidth]{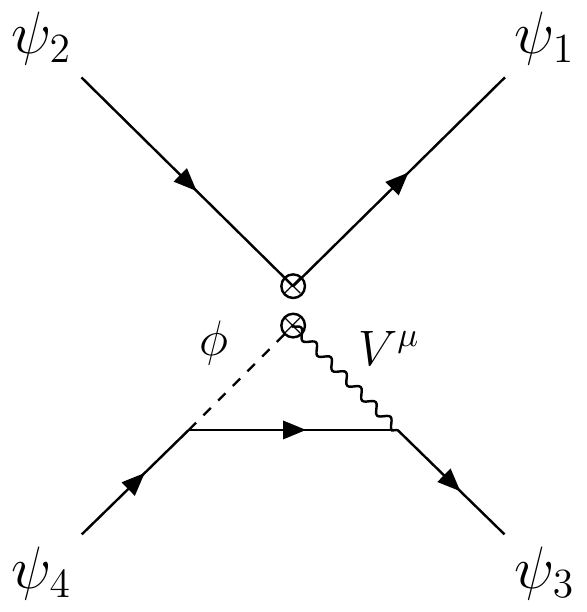}
		\caption{}
		\label{fig:4f_HA_2}
	\end{subfigure}
	\caption{The seven  structures that contribute to the four-fermion operator anomalous dimension matrix at the one-loop level. The $\psi_i$ are operator-dependent external fermions. The fermion inside the loop is related to $\psi_i$ through Yukawa or gauge couplings.}
	\label{fig:feyn}
\end{figure}

\subsection{Counterterms}

\begin{figure}[t]
	\centering
	\begin{subfigure}{0.24\textwidth}
		\includegraphics[width=\textwidth]{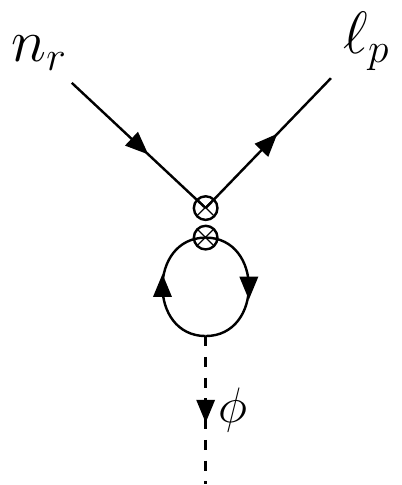}
		\label{fig:Xi_n}
	\end{subfigure}
	\begin{subfigure}{0.24\textwidth}
		\includegraphics[width=\textwidth]{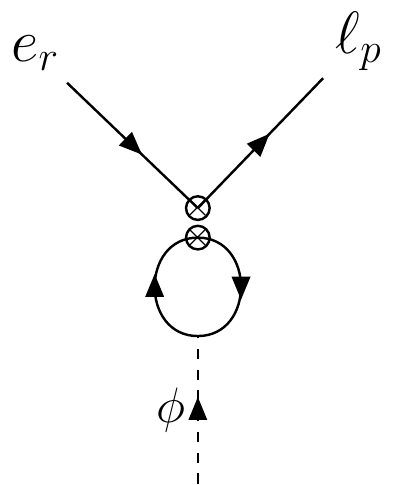}
		\label{fig:Xi_e}
	\end{subfigure}
	\begin{subfigure}{0.24\textwidth}
		\includegraphics[width=\textwidth]{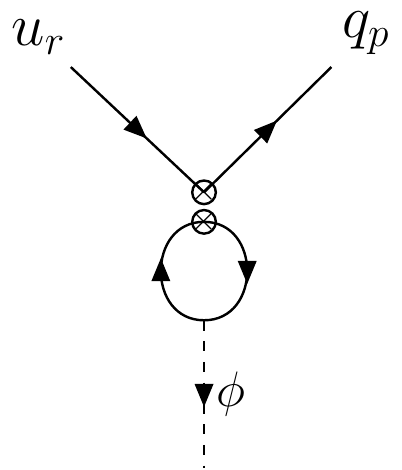}
		\label{fig:Xi_u}
	\end{subfigure}
	\begin{subfigure}{0.24\textwidth}
		\includegraphics[width=\textwidth]{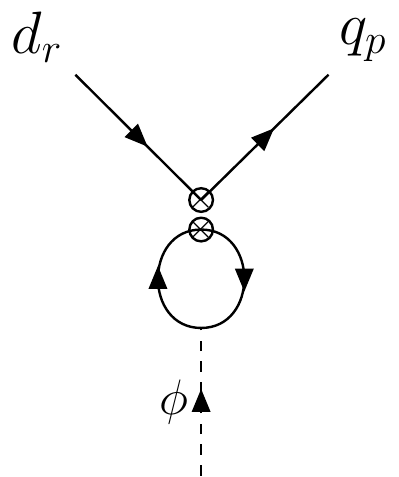}
		\label{fig:Xi_d}
	\end{subfigure}
	\caption{The Feynman diagrams associated with the $\xi$ parameters in Eq.~(\ref{eq:xi}).}
	\label{fig:Xi}
\end{figure}

The corrections from counterterms cancel the ultraviolet (UV) divergence from the  one-loop diagrams. In the one-loop diagrams, there are 14 different structures as in Fig.~\ref{fig:feyn}; there are seven counterparts to those shown. We display the UV divergent part of each structure in Fig.~\ref{fig:feyn}.
 The UV divergent parts in Figs.~\ref{fig:4f_12} to~\ref{fig:4f_14} are of the form
\beqa
\mathcal{D}_a & = & -\frac{1}{64\pi^2\epsilon} (\bar\psi_1 \gamma^\mu \Gamma_1 \gamma_\mu \psi_2)  (\psi_3 \Gamma_2 \psi_4)\,,\\
\mathcal{D}_b & = & -\frac{1}{64\pi^2\epsilon} (\bar\psi_1 \gamma^\mu \Gamma_1 \psi_2) (\bar\psi_3 \gamma_\mu  \Gamma_2 \psi_4)\,,\\
\mathcal{D}_c & = & -\frac{1}{64\pi^2\epsilon} ( \bar \psi_1\Gamma_1 \gamma^\mu \psi_2) (\bar\psi_3\gamma_\mu  \Gamma_2\psi_4)\,,
\eeqa
where $\Gamma_1$ and $\Gamma_2$ are the Lorentz structures for the upper and lower vertex, respectively. In Fig.~\ref{fig:4f_Xi}, $\Gamma_1 $ has to be $P_2$, which is the projection operator of the chiral fermion field $\psi_2$, because for the other possibilities, the UV divergent parts vanish. Thus we obtain
\beq
\mathcal{D}_d = \frac{1}{16\pi^2\epsilon} (\bar\psi_1 P_2 \psi_2) (\bar\psi_3 P_4 \psi_4)\,. 
\eeq
The UV divergent part of Fig.~\ref{fig:4f_Hf} is of the form
\beq
\mathcal{D}_e = -\frac{1}{32\pi^2\epsilon} (\bar\psi_1 \gamma^\mu P_2 \psi_2) (\bar\psi_3 \gamma_\mu  P_4 \psi_4)\,.
\eeq
For the dipole operators in Figs.~\ref{fig:4f_HA} and~\ref{fig:4f_HA_2}, the UV divergent parts are of the form
\beqa
\mathcal{D}_f &=&  \frac{i}{64\pi^2\epsilon} (\bar\psi_1 \sigma^{\mu\nu} P_2 \psi_2) (\bar\psi_3 \gamma^\beta \gamma^\alpha  P_4 \psi_4) (g^{\mu\beta} g^{\nu\alpha} - g^{\mu\alpha} g^{\nu\beta}) \nonumber\\
 &=& \frac{1}{32\pi^2\epsilon} (\bar\psi_1 \sigma^{\mu\nu} P_2 \psi_2) (\bar\psi_3  \sigma_{\mu\nu} P_4 \psi_4)\,,\\
\mathcal{D}_g &=&  \frac{i}{64\pi^2\epsilon} (\bar\psi_1 \sigma^{\mu\nu} P_2 \psi_2) (\bar\psi_3 \gamma^\alpha \gamma^\beta  P_4 \psi_4) (g^{\mu\beta} g^{\nu\alpha} - g^{\mu\alpha} g^{\nu\beta}) \nonumber\\
&=&  -\frac{1}{32\pi^2\epsilon} (\bar\psi_1 \sigma^{\mu\nu} P_2 \psi_2) (\bar\psi_3  \sigma_{\mu\nu} P_4 \psi_4)\,.
\eeqa
 To simplify our results further, we follow Ref.~\cite{Jenkins:2013wua} and define the 
 amplitudes in Fig.~\ref{fig:Xi} in connection with Fig.~\ref{fig:4f_Xi}:
\beqa
\xi_{\us{pr}{n}} & = & 2\C_{\us{pwvr}{\ell n}} [Y^\dagger_n]_{wv} - N_c\C^{(1)}_{\us{prvw}{\ell nqd}} [Y_d]_{wv}  -N_c  \C_{\us{prvw}{\ell nuq}} [Y^\dagger_u]_{wv} - \C_{\us{prvw}{\ell n\ell e}} [Y_e]_{wv}\,,\nonumber\\
\xi_{\us{pr}{e}} & = & 2\C_{\us{pwvr}{\ell e}} [Y^\dagger_e]_{wv} - N_c\C_{\us{prvw}{\ell edq}} [Y^\dagger_d]_{wv} + N_c  \C^{(1)}_{\us{prvw}{\ell equ}} [Y_u]_{wv} - \C_{\us{vwpr}{\ell n\ell e}} [Y_n]_{wv}\,,\nonumber\\
\xi_{\us{pr}{u}}& = & 2(\C^{(1)}_{\us{pwvr}{qu}}+C_{F,3}\C^{(8)}_{\us{pwvr}{qu}}) [Y^\dagger_u]_{wv} - (N_c\C^{(1)}_{\us{prvw}{quqd}} + \frac{1}{2}\C^{(1)}_{\us{vrpw}{quqd}} + \frac{1}{2} C_{F,3}\C^{(8)}_{\us{vrpw}{quqd}}) [Y^\dagger_d]_{wv} + \C^{(1)}_{\us{vwpr}{\ell equ}}[Y_e]_{wv}-\C^*_{\us{vwrp}{\ell nuq}}[Y_n^\dagger]_{vw}\,,\nonumber\\
\xi_{\us{pr}{d}} & = & 2(\C^{(1)}_{\us{pwvr}{qd}}+C_{F,3}\C^{(8)}_{\us{pwvr}{qd}}) [Y^\dagger_d]_{wv} - (N_c\C^{(1)}_{\us{vrpw}{quqd}} + \frac{1}{2}\C^{(1)}_{\us{prvw}{quqd}} + \frac{1}{2} C_{F,3}\C^{(8)}_{\us{prvw}{quqd}}) [Y^\dagger_u]_{wv} - \C^{*}_{\us{vwrp}{\ell edq}}[Y^\dagger_e]_{vw}-\C^{(1)}_{\us{vwpr}{\ell nqd}}[Y_n]_{wv}\,,\nonumber\\
\label{eq:xi}
\eeqa
where the quadratic Casimir $C_{F,3} = \frac{4}{3}$ and the number of colors $N_c = 3$.
The $\xi$ parameter for right-handed neutrinos $n$ ($\xi_n$), corresponds to the new terms in SMNEFT, while the  last terms in $\xi_e$, $\xi_u$ and $\xi_d$ are contributions from the right-handed neutrino Yukawa couplings not present in SMEFT.

\section{Results}

In this section, we present the Yukawa coupling contributions to the one-loop RGE for all four-fermion SMNEFT operators, and the new RGE terms for the four-fermion SMEFT operators due to the mixing between SMEFT and SMNEFT  operators via the right-handed neutrino Yukawa couplings $Y_n$. The contributions from the fermionic operators come from the Feynman diagrams in Figs.~\ref{fig:4f_12} to~\ref{fig:4f_Xi}, with contributions from Fig.~\ref{fig:4f_Xi} given by the $\xi$ parameters. 

\begin{boldmath}
	\subsection{Anomalous dimensions from Yukawa couplings: SMNEFT}
\end{boldmath}

The bosonic operators in Table~\ref{Tb:SMNEFT} contribute to the SMNEFT ADM but not the SMEFT ADM. 
The contribution from the bosonic operator 
$\psi^2\phi^2 D$ is shown in Fig.~\ref{fig:4f_Hf}. The RGE of  the dipole operators $\Q_{\ell n\ell e}$ and $\Q^{(3)}_{\ell nqd}$  is modified by the $\psi^2 X \phi$ 
operators in Table~\ref{Tb:SMNEFT} and the relevant diagrams are shown in Figs.~\ref{fig:4f_HA} and~\ref{fig:4f_HA_2}. These terms contain both gauge and Yukawa coupling contributions.

\newpage
\begin{boldmath}
	\subsubsection{$(\bar RR)(\bar RR)$}
\end{boldmath}

\beqa
\dot \C_{\underset{prst}{nd}} &=& -2[Y_n Y_n^\dagger]_{pr} \C_{\underset{st}{\phi d}}+2[Y_d Y_d^\dagger]_{st} \C_{\underset{pr}{\phi n}}-2[Y_n]_{pv}[Y_n^\dagger]_{wr}\C_{\underset{vwst}{\ell d}}-2[Y_d]_{sv}[Y_d^\dagger]_{wt}\C_{\underset{vwpr}{qn}} \nonumber\\
&\quad& - ([Y_n]_{pv}[Y_d]_{sw} \C^{(1)}_{\underset{vrwt}{\ell nqd}} + [Y_n^\dagger]_{vr}[Y_d^\dagger]_{wt}  \C^{(1)*}_{\underset{vpws}{\ell nqd}}) + 12 ([Y_n]_{pv}[Y_d]_{sw} \C^{(3)}_{\underset{vrwt}{\ell nqd}} + [Y_n^\dagger]_{vr}[Y_d^\dagger]_{wt}  \C^{(3)*}_{\underset{vpws}{\ell nqd}})\nonumber\\
&\quad&+\gamma^{(Y)}_{\underset{pv}{n}} \C_{\underset{vrst}{nd}}+\gamma^{(Y)}_{\underset{sv}{d}} \C_{\underset{prvt}{nd}}+ \C_{\underset{pvst}{nd}} \gamma^{(Y)}_{\underset{vr}{n}}+ \C_{\underset{prsv}{nd}} \gamma^{(Y)}_{\underset{vt}{d}}\,,
\eeqa
%
\beqa
\dot \C_{\underset{prst}{nu}} &=& -2[Y_n Y_n^\dagger]_{pr} \C_{\underset{st}{\phi u}}-2[Y_u Y_u^\dagger]_{st} \C_{\underset{pr}{\phi n}}-2[Y_n]_{pv}[Y_n^\dagger]_{wr}\C_{\underset{vwst}{\ell u}}-2[Y_u]_{sv}[Y_u^\dagger]_{wt}\C_{\underset{vwpr}{qn}}  \nonumber\\ &\quad&+([Y_n]_{pv}[Y_u^\dagger]_{wt} \C_{\underset{vrsw}{\ell nuq}}+[Y^\dagger_n]_{vr}[Y_u]_{sw} \C^*_{\underset{vptw}{\ell nuq}})\nonumber\\
&\quad&+\gamma^{(Y)}_{\underset{pv}{n}} \C_{\underset{vrst}{nu}}+\gamma^{(Y)}_{\underset{sv}{u}} \C_{\underset{prvt}{nu}}+ \C_{\underset{pvst}{nu}} \gamma^{(Y)}_{\underset{vr}{n}}+ \C_{\underset{prsv}{nu}} \gamma^{(Y)}_{\underset{vt}{u}}\,,
\eeqa
\beqa
\dot \C_{\underset{prst}{ne}} &=& -2[Y_n Y_n^\dagger]_{pr} \C_{\underset{st}{\phi e}}+2[Y_e Y_e^\dagger]_{st} \C_{\underset{pr}{\phi n}}+2[Y_e Y_n^\dagger]_{sr} \C_{\underset{pt}{\phi ne}}+2[Y_n Y_e^\dagger]_{pt} \C^*_{\underset{rs}{\phi ne}}-2[Y_n]_{pv}[Y_n^\dagger]_{wr}\C_{\underset{vwst}{\ell e}} \nonumber\\ 
&\quad& -([Y_n]_{pv}[Y_e]_{sw} \C_{\underset{vrwt}{\ell n\ell e}}+[Y^\dagger_n]_{vr}[Y_e^\dagger]_{wt} \C^*_{\underset{vpws}{\ell n\ell e}})+ ([Y_n]_{pw}[Y_e]_{sv} \C_{\underset{vrwt}{\ell n\ell e}}+[Y^\dagger_n]_{wr}[Y_e^\dagger]_{vt} \C^*_{\underset{vpws}{\ell n\ell e}})\nonumber\\
&\quad&-2[Y_e]_{sv}[Y_e^\dagger]_{wt}\C_{\underset{vwpr}{\ell n}}+\gamma^{(Y)}_{\underset{pv}{n}} \C_{\underset{vrst}{ne}}+\gamma^{(Y)}_{\underset{sv}{e}} \C_{\underset{prvt}{ne}}+ \C_{\underset{pvst}{ne}} \gamma^{(Y)}_{\underset{vr}{n}}+ \C_{\underset{prsv}{ne}} \gamma^{(Y)}_{\underset{vt}{e}}\,,
\eeqa
\beqa
\dot \C_{\underset{prst}{nn}} &=& -[Y_n Y_n^\dagger]_{pr} \C_{\underset{st}{\phi n}}-[Y_n Y_n^\dagger]_{st} \C_{\underset{pr}{\phi n}}-[Y_n]_{pv}[Y_n^\dagger]_{wr}\C_{\underset{vwst}{\ell n}}-[Y_n]_{sv}[Y_n^\dagger]_{wt}\C_{\underset{vwpr}{\ell n}}  \nonumber\\
&\quad&+\gamma^{(Y)}_{\underset{pv}{n}} \C_{\underset{vrst}{nn}}+\gamma^{(Y)}_{\underset{sv}{n}} \C_{\underset{prvt}{nn}}+ \C_{\underset{pvst}{nn}} \gamma^{(Y)}_{\underset{vr}{n}}+ \C_{\underset{prsv}{nn}} \gamma^{(Y)}_{\underset{vt}{n}}\,,
\eeqa
\beqa
\dot \C_{\underset{prst}{nedu}} &=& 2[Y_d Y_u^\dagger]_{st} \C_{\underset{pr}{\phi ne}}+2[Y_n Y_e^\dagger]_{pr} \C^*_{\underset{ts}{\phi ud}} - [Y_n]_{pv}[Y_d]_{sw} (\C^{(1)}_{\underset{vrwt}{\ell equ}} - 12 \C^{(3)}_{\underset{vrwt}{\ell equ}}) + [Y_n]_{pv}[Y^\dagger_u]_{wt} \C_{\underset{vrsw}{\ell edq}} \nonumber\\
&\quad& + [Y^\dagger_e]_{vr}[Y^\dagger_u]_{wt} (\C^{(1)*}_{\underset{vpws}{\ell nqd}} - 12 \C^{(3)*}_{\underset{vpws}{\ell nqd}})+ [Y^\dagger_e]_{vr}[Y_d]_{sw} \C^*_{\underset{vptw}{\ell nuq}} \nonumber\\
&\quad&+\gamma^{(Y)}_{\underset{pv}{n}} \C_{\underset{vrst}{nedu}}+\gamma^{(Y)}_{\underset{sv}{d}} \C_{\underset{prvt}{nedu}}+ \C_{\underset{pvst}{nedu}} \gamma^{(Y)}_{\underset{vr}{e}}+ \C_{\underset{prsv}{nedu}} \gamma^{(Y)}_{\underset{vt}{u}}\,.
\eeqa

\begin{boldmath}
	\subsubsection{$(\bar LL)(\bar RR)$}
\end{boldmath}

\beqa
\dot \C_{\underset{prst}{qn}} &=& [Y_u^\dagger Y_u - Y_d^\dagger Y_d]_{pr} \C_{\underset{st}{\phi n}} - 2[Y_n Y_n^\dagger]_{st} \C^{(1)}_{\underset{pr}{\phi q}}-2[Y_n]_{sv}[Y_n^\dagger]_{wt}\C^{(1)}_{\underset{vwpr}{\ell q}} -[Y_u]_{wr}[Y_u^\dagger]_{pv}\C_{\underset{stvw}{nu}}-[Y_d]_{wr}[Y_d^\dagger]_{pv}\C_{\underset{stvw}{nd}}\nonumber\\
&\quad& +\frac{1}{2}([Y_n]_{sw}[Y_d]_{vr}C^{(1)}_{\us{wtpv}{\ell nqd}} + [Y^\dagger_n]_{wt}[Y^\dagger_d]_{pv}C^{(1)*}_{\us{wsrv}{\ell nqd}} )+6([Y_n]_{sw}[Y_d]_{vr}C^{(3)}_{\us{wtpv}{\ell nqd}} + [Y^\dagger_n]_{wt}[Y^\dagger_d]_{pv}C^{(3)*}_{\us{wsrv}{\ell nqd}} ) \nonumber\\
&\quad& -\frac{1}{2}([Y_n]_{sw}[Y^\dagger_u]_{pv}C_{\us{wtvr}{\ell nuq}} + [Y^\dagger_n]_{wt}[Y_u]_{vr}C^*_{\us{wsvp}{\ell nuq}} ) \nonumber\\
&\quad&+\gamma^{(Y)}_{\underset{pv}{q}} \C_{\underset{vrst}{qn}}+\gamma^{(Y)}_{\underset{sv}{n}} \C_{\underset{prvt}{qn}}+ \C_{\underset{pvst}{qn}} \gamma^{(Y)}_{\underset{vr}{q}}+ \C_{\underset{prsv}{qn}} \gamma^{(Y)}_{\underset{vt}{n}}\,,
\eeqa
\newpage
\beqa
\dot \C_{\underset{prst}{\ell n}} &=& [Y_n^\dagger Y_n - Y_e^\dagger Y_e]_{pr} \C_{\underset{st}{\phi n}}-2[Y_n Y_n^\dagger]_{st} \C^{(1)}_{\underset{pr}{\phi\ell}} + [Y^\dagger_n]_{pw} [Y_n]_{sv} \C_{\underset{vrwt}{\ell n}} + [Y^\dagger_n]_{vt} [Y_n]_{wr} \C_{\underset{pvsw}{\ell n}} - [Y^\dagger_e]_{pv} [Y_e]_{wr} \C_{\underset{stvw}{ne}} \nonumber\\
&\quad&- 2 [Y^\dagger_n]_{pv}[Y_n]_{wr} \C_{\underset{vtsw}{nn}} - 2[Y^\dagger_n]_{pv}[Y_n]_{wr} \C_{\underset{vwst}{nn}}
-  2 [Y^\dagger_n]_{vt}[Y_n]_{sw} \C_{\underset{pvwr}{\ell\ell}}- 4[Y^\dagger_n]_{wt}[Y_n]_{sv} \C_{\underset{prvw}{\ell\ell}}\nonumber\\
&\quad&+ \frac{1}{2}([Y_e]_{wr}[Y_n]_{sv} \C_{\underset{vtpw}{\ell n\ell e}} + [Y^\dagger_e]_{pv}[Y^\dagger_n]_{wt} \C^*_{\underset{wsrv}{\ell n\ell e}}) + ([Y_e]_{wr}[Y_n]_{sv} \C_{\underset{ptvw}{\ell n\ell e}} + [Y^\dagger_e]_{pv}[Y^\dagger_n]_{wt} \C^*_{\underset{rswv}{\ell n\ell e}}) \nonumber\\
&\quad&+ [Y_n]_{sr} \xi_{\us{pt}{n}} + [Y^\dagger_n]_{pt} \xi^*_{\us{rs}{n}}+\gamma^{(Y)}_{\underset{pv}{\ell}} \C_{\underset{vrst}{\ell n}}+\gamma^{(Y)}_{\underset{sv}{n}} \C_{\underset{prvt}{\ell n}}+ \C_{\underset{pvst}{\ell n}} \gamma^{(Y)}_{\underset{vr}{\ell}}+ \C_{\underset{prsv}{\ell n}} \gamma^{(Y)}_{\underset{vt}{n}}\,.
\eeqa

\begin{boldmath}
	\subsubsection{$(\bar LR)(\bar RL)$ and $(\bar LR)(\bar LR)$}
\end{boldmath}

\beqa
\dot \C_{\underset{prst}{\ell n\ell e}} &=& -4 ([Y^\dagger_n]_{vr}[Y^\dagger_e]_{wt} \C_{\underset{pvsw}{\ell\ell}} - [Y^\dagger_n]_{vr}[Y^\dagger_e]_{wt} \C_{\underset{svpw}{\ell\ell}}) + 4 ([Y^\dagger_n]_{wr}[Y^\dagger_e]_{vt} \C_{\underset{pvsw}{\ell\ell}} - [Y^\dagger_n]_{wt}[Y^\dagger_e]_{vt} \C_{\underset{svpw}{\ell\ell}}) \nonumber\\
&\quad&- 4([Y^\dagger_n]_{pv}[Y^\dagger_e]_{sw} \C_{\underset{vrwt}{ne}} - [Y^\dagger_n]_{sv}[Y^\dagger_e]_{pw} \C_{\underset{vrwt}{ne}} )+ 4 [Y^\dagger_n]_{sw}[Y^\dagger_e]_{vt} \C_{\us{pvwr}{\ell n}} + 4 [Y^\dagger_n]_{vr}[Y^\dagger_e]_{pw} \C_{\underset{svwt}{\ell e}} \nonumber\\
&\quad&+4 g_1(y_e+y_\ell)\C_{\us{pr}{nB}}[Y_e^\dagger]_{st}-8 g_1(y_e+y_\ell)\C_{\us{sr}{nB}}[Y_e^\dagger]_{pt} - 6 g_2\C_{\us{pr}{nW}}[Y_e^\dagger]_{st} + 12g_2\C_{\us{sr}{nW}}[Y_e^\dagger]_{pt}\nonumber\\
&\quad&+4 g_1(y_n+y_\ell)\C_{\us{st}{eB}}[Y_n^\dagger]_{pr}-8 g_1(y_n+y_\ell)\C_{\us{pt}{eB}}[Y_n^\dagger]_{sr}- 6 g_2\C_{\us{st}{eW}}[Y_n^\dagger]_{pr} + 12g_2\C_{\us{pt}{eW}}[Y_n^\dagger]_{sr}\nonumber\\
&\quad&- 2 \xi_{\us{pr}{n}} [Y^\dagger_e]_{st} - 2 \xi_{\us{st}{e}} [Y^\dagger_n]_{pr}+\gamma^{(Y)}_{\underset{pv}{\ell}} \C_{\underset{vrst}{\ell n\ell e}}+\gamma^{(Y)}_{\underset{sv}{\ell}} \C_{\underset{prvt}{\ell n\ell e}}+ \C_{\underset{pvst}{\ell n\ell e}} \gamma^{(Y)}_{\underset{vr}{n}}+ \C_{\underset{prsv}{\ell n\ell e}} \gamma^{(Y)}_{\underset{vt}{e}}\,,
\eeqa
\beqa
\dot \C^{(1)}_{\underset{prst}{\ell nqd}} &=& -2[Y^\dagger_n]_{vr} [Y^\dagger_e]_{pw} \C^*_{\underset{vwts}{\ell edq}} + 2[Y^\dagger_n]_{pw} [Y^\dagger_d]_{vt} \C_{\underset{svwr}{qn}}+ 2[Y^\dagger_e]_{pw} [Y^\dagger_u]_{sv} \C^*_{\underset{rwtv}{nedu}} + 2[Y^\dagger_n]_{vr} [Y^\dagger_d]_{sw} \C_{\underset{pvwt}{\ell d}}\nonumber\\
&\quad& - 2[Y^\dagger_d]_{wt} [Y^\dagger_u]_{sv} \C_{\underset{prvw}{\ell nuq}} - 2[Y^\dagger_n]_{pw} [Y^\dagger_d]_{sv} \C_{\underset{wrvt}{nd}} - 2[Y^\dagger_n]_{vr} [Y^\dagger_d]_{wt} \C^{(1)}_{\underset{pvsw}{\ell q}} + 6[Y^\dagger_n]_{vr} [Y^\dagger_d]_{wt} \C^{(3)}_{\underset{pvsw}{\ell q}} \nonumber\\
&\quad&- 2 \xi_{\us{pr}{n}} [Y_d^\dagger]_{st} - 2 \xi_{\us{st}{d}} [Y_n^\dagger]_{pr} +\gamma^{(Y)}_{\underset{pv}{\ell}} \C^{(1)}_{\underset{vrst}{\ell nqd}}+\gamma^{(Y)}_{\underset{sv}{q}} \C^{(1)}_{\underset{prvt}{\ell nqd}}+ \C^{(1)}_{\underset{pvst}{\ell nqd}} \gamma^{(Y)}_{\underset{vr}{n}}+ \C^{(1)}_{\underset{prsv}{\ell nqd}} \gamma^{(Y)}_{\underset{vt}{d}}\,,
\eeqa
\beqa
\dot \C^{(3)}_{\underset{prst}{\ell nqd}} &=& - \frac{1}{2}[Y^\dagger_e]_{pw} [Y^\dagger_u]_{sv} \C^*_{\underset{rwtv}{nedu}} + \frac{1}{2}[Y^\dagger_n]_{vr} [Y^\dagger_d]_{wt} \C^{(1)}_{\underset{pvsw}{\ell q}} - \frac{3}{2}[Y^\dagger_n]_{vr} [Y^\dagger_d]_{wt} \C^{(3)}_{\underset{pvsw}{\ell q}} \nonumber\\
&\quad&  \frac{1}{2}[Y^\dagger_n]_{vr} [Y^\dagger_d]_{sw} \C_{\underset{pvwt}{\ell d}} + \frac{1}{2}[Y^\dagger_n]_{pw} [Y^\dagger_d]_{vt} \C_{\underset{svwr}{qn}}  + \frac{1}{2}[Y^\dagger_n]_{pw} [Y^\dagger_d]_{sv} \C_{\underset{wrvt}{nd}} \nonumber\\
&\quad&-g_1(y_d+y_q)\C_{\us{pr}{nB}}[Y_d^\dagger]_{st}-g_1(y_n+y_\ell)\C_{\us{st}{dB}}[Y_n^\dagger]_{pr} + \frac{3}{2}g_2 \C_{\us{pr}{nW}}[Y_d^\dagger]_{st}+ \frac{3}{2}g_2 \C_{\us{st}{dW}}[Y_n^\dagger]_{pr}\nonumber\\
&\quad&+\gamma^{(Y)}_{\underset{pv}{\ell}} \C^{(3)}_{\underset{vrst}{\ell nqd}}+\gamma^{(Y)}_{\underset{sv}{q}} \C^{(3)}_{\underset{prvt}{\ell nqd}}+ \C^{(3)}_{\underset{pvst}{\ell nqd}} \gamma^{(Y)}_{\underset{vr}{n}}+ \C^{(3)}_{\underset{prsv}{\ell nqd}} \gamma^{(Y)}_{\underset{vt}{d}}\,,
\eeqa
\beqa
\dot \C_{\underset{prst}{\ell nuq}} &=& 2[Y^\dagger_n]_{wr} [Y^\dagger_e]_{pv} \C^{(1) *}_{\underset{wvts}{\ell equ}} - 2[Y^\dagger_n]_{pw} [Y_u]_{sv} \C_{\underset{vtwr}{qn}} + 2[Y^\dagger_n]_{pw} [Y_u]_{vt} \C_{\underset{wrsv}{nu}}+ 2[Y^\dagger_e]_{pw} [Y_d]_{vt} \C^*_{\underset{rwvs}{nedu}}\nonumber\\
&\quad& +2[Y^\dagger_n]_{vr} [Y_u]_{wt} \C^{(1)}_{\underset{pvwt}{\ell q}}+6[Y^\dagger_n]_{vr} [Y_u]_{wt} \C^{(3)}_{\underset{pvwt}{\ell q}} - 2[Y^\dagger_n]_{vr} [Y_u]_{wt} \C_{\underset{pvsw}{\ell u}} - 2 [Y^\dagger_u]_{sv} [Y_d]_{wt} \C^{(1)}_{\underset{prvw}{\ell nqd}}\nonumber\\
&\quad& - 2 \xi_{\us{pr}{n}} [Y_u]_{st} - 2 \xi_{\us{st}{u}}^* [Y_n^\dagger]_{pr} +\gamma^{(Y)}_{\underset{pv}{\ell}} \C_{\underset{vrst}{\ell nuq}}+\gamma^{(Y)}_{\underset{sv}{u}} \C_{\underset{prvt}{\ell nuq}}+ \C_{\underset{pvst}{\ell nuq}} \gamma^{(Y)}_{\underset{vr}{n}}+ \C_{\underset{prsv}{\ell nuq}} \gamma^{(Y)}_{\underset{vt}{q}}\,,
\eeqa
where $y_n = 0$, $y_e = -1$, $y_{\ell} = -1/2$, $y_d = - 1/3$, $y_u = 2/3$, and $y_q = 1/6$ are the hypercharges.
\subsection{Anomalous dimensions from Yukawa couplings: SMEFT}
The Yukawa interactions of the right-handed neutrinos modify the RGE of the four-fermion SMEFT operators  listed in Table~\ref{Tb:SMEFT}. We only provide the additional terms induced by the right-handed neutrino Yukawa couplings $Y_n$. For the operators in the lower panel of Table~\ref{Tb:SMEFT}, the anomalous dimensions are modified via the $\xi$ parameters in Eq.~(\ref{eq:xi}).
\begin{table}
	\centering
	\renewcommand{\arraystretch}{1.5}
	\begin{tabular}{|c|c|c|c|c|c|}
		\hline
		\multicolumn{2}{|c|}{$(\bar LL)(\bar LL)$} &
		\multicolumn{2}{|c|}{$(\bar LL)(\bar RR)$} &
		\multicolumn{2}{|c|}{$(\bar LR)(\bar RL)$ and $(\bar LR)(\bar LR)$ }\\
		\hline
		$\Q_{\ell\ell}$  & $(\bar \ell_p \gamma_\mu \ell_r)(\bar \ell_s \gamma^\mu \ell_t)$ &
		$\Q_{\ell d}$  & $(\bar \ell_p \gamma_\mu \ell_r)(\bar d_s \gamma^\mu d_t)$ &
		$\Q^{(1)}_{\ell equ}$ & $(\bar \ell^j_p  e_r)\epsilon_{jk}(\bar q^k_s u_t)$   \\
		$\Q^{(1)}_{\ell q}$  & $(\bar \ell_p \gamma_\mu \ell_r)(\bar q_s \gamma^\mu q_t)$ &
		$\Q_{\ell u}$ & $(\bar \ell_p \gamma_\mu \ell_r)(\bar u_s\gamma^\mu u_t)$ &
		$\Q^{(3)}_{\ell equ}$ &  $(\bar \ell^j_p  \sigma_{\mu\nu} e_r)\epsilon_{jk}(\bar q^k_s \sigma^{\mu\nu} u_t)$  \\
		$\Q^{(3)}_{\ell q}$ & $(\bar \ell_p \gamma_\mu \tau^I \ell_r)(\bar q_s \gamma^\mu \tau^I q_t)$ & $\Q_{\ell e}$ 
		& $(\bar \ell_p \gamma_\mu \ell_r)(\bar e_s \gamma^\mu e_t)$ &
		$\Q_{\ell e d q}$ & $(\bar \ell^j_p  e_r)(\bar d_s q^j_t)$      \\
		\hline
		\hline
		  &  & $\Q^{(1)}_{qu}$ & $(\bar q_p \gamma_\mu q_r)(\bar u_s \gamma^\mu u_t)$  &
		$\Q^{(1)}_{quqd}$  & $(\bar \ell^j_p  n_r)(\bar u_s q^j_t)$  \\
		    &    & $\Q^{(8)}_{qu}$ & $(\bar q_p \gamma_\mu T^A q_r)(\bar u_s \gamma^\mu T^A u_t)$ &	&   \\
		   &  & $\Q^{(1)}_{qd}$ & $(\bar q_p \gamma_\mu  q_r)(\bar d_s \gamma^\mu d_t)$ &	& \\
		    &   & $\Q^{(8)}_{qd}$ & $(\bar q_p \gamma_\mu T^A q_r)(\bar d_s \gamma^\mu T^A d_t)$ &	&\\
		\hline
	\end{tabular}
	\bigskip
	\captionsetup{width=0.9\textwidth}
	\caption{\small \label{tab:SMEFTops} The 14 four-fermion SMEFT operators whose anomalous dimensions are modified by right-handed neutrino Yukawa couplings. Here, $I (A)$ is the adjoint index of $SU(2)_L$ ($SU(3)_C$). }
	\label{Tb:SMEFT}
\end{table}

\begin{boldmath}
	\subsubsection{$(\bar LL)(\bar LL)$}
\end{boldmath}

\beqa
\dot \C_{\underset{prst}{\ell \ell}} &\supset&\frac{1}{2}[Y_n^\dagger Y_n]_{pr} (\C^{(1)}_{\underset{st}{\phi\ell}}+\C^{(3)}_{\underset{st}{\phi\ell}})+\frac{1}{2}[Y_n^\dagger Y_n]_{st} (\C^{(1)}_{\underset{pr}{\phi\ell}}+\C^{(3)}_{\underset{pr}{\phi\ell}})-\frac{1}{2} [Y_n]_{sv}[Y_n]_{wt} \C_{\us{prvw}{\ell n}} -\frac{1}{2} [Y_n]_{pv}[Y_n]_{wr} \C_{\us{stvw}{\ell n}} \nonumber\\
&\quad& - \frac{1}{2} ([Y_n]_{vr}[Y_e]_{wt} \C_{\us{pvsw}{\ell n\ell e}} + [Y_n]_{wt}[Y_e]_{vr} \C_{\us{swpv}{\ell n\ell e}}) - \frac{1}{2} ([Y^\dagger_n]_{pv}[Y^\dagger_e]_{sw} \C^*_{\us{rvtw}{\ell n\ell e}} + [Y^\dagger_n]_{sw}[Y^\dagger_e]_{pv} \C^*_{\us{twrv}{\ell n\ell e}})\,,\nonumber\\
\eeqa
%
\beqa
\dot \C^{(1)}_{\underset{prst}{\ell q}} &\supset& [Y_n^\dagger Y_n]_{pr} \C^{(1)}_{\underset{st}{\phi q}} - [Y^\dagger_n]_{pv}[Y_n]_{wr} \C_{\underset{stvw}{qn}} + \frac{1}{4} ([Y_n]_{vr}[Y_u]_{sw} \C_{\us{pvwt}{\ell nuq}} + [Y^\dagger_n]_{pv}[Y^\dagger_u]_{wt} \C^*_{\us{rvws}{\ell nuq}} )\nonumber\\
&\quad&-\frac{1}{4}  ([Y_n]_{vr}[Y_d]_{wt} \C^{(1)}_{\us{pvsw}{\ell nqd}} + [Y^\dagger_n]_{pv}[Y^\dagger_d]_{sw} \C^{(1)*}_{\us{rvtw}{\ell nqd}} ) + 3 ([Y_n]_{vr}[Y_d]_{wt} \C^{(3)}_{\us{pvsw}{\ell nqd}} + [Y^\dagger_n]_{pv}[Y^\dagger_d]_{sw} \C^{(3)*}_{\us{rvtw}{\ell nqd}})\,, \nonumber\\
\eeqa
\beqa
\dot \C^{(3)}_{\underset{prst}{\ell q}} &\supset& -[Y_n^\dagger Y_n]_{pr} \C^{(3)}_{\underset{st}{\phi q}} + \frac{1}{4} ([Y_n]_{vr}[Y_u]_{sw} \C_{\us{pvwt}{\ell nuq}} + [Y^\dagger_n]_{pv}[Y^\dagger_u]_{wt} \C^*_{\us{rvws}{\ell nuq}} )\nonumber\\
&\quad&+\frac{1}{4}  ([Y_n]_{vr}[Y_d]_{wt} \C^{(1)}_{\us{pvsw}{\ell nqd}} + [Y^\dagger_n]_{pv}[Y^\dagger_d]_{sw} \C^{(1)*}_{\us{rvtw}{\ell nqd}} )-3 ([Y_n]_{vr}[Y_d]_{wt} \C^{(3)}_{\us{pvsw}{\ell nqd}} + [Y^\dagger_n]_{pv}[Y^\dagger_d]_{sw} \C^{(3)*}_{\us{rvtw}{\ell nqd}} )\,. \nonumber\\
\eeqa

\begin{boldmath}
	\subsubsection{$(\bar LL)(\bar RR)$}
\end{boldmath}

\beqa
\dot \C_{\underset{prst}{\ell d}} &\supset& [Y_n^\dagger Y_n]_{pr} \C_{\underset{st}{\phi d}} - [Y_n]_{pv}[Y_n^\dagger]_{wr} \C_{\underset{vwst}{nd}} + \frac{1}{2}([Y_d]_{sw}[Y_n]_{vr} \C^{(1)}_{\us{pvwt}{\ell nqd}} + [Y^\dagger_d]_{wt}[Y^\dagger_n]_{pv} \C^{(1)*}_{\us{rvws}{\ell nqd}}  )\nonumber\\
&\quad&+ 6([Y_d]_{sw}[Y_n]_{vr} \C^{(3)}_{\us{pvwt}{\ell nqd}} + [Y^\dagger_d]_{wt}[Y^\dagger_n]_{pv} \C^{(3)*}_{\us{rvws}{\ell nqd}})\,,
\eeqa
\beqa
\dot \C_{\underset{prst}{\ell u}} &\supset& [Y_n^\dagger Y_n]_{pr} \C_{\underset{st}{\phi u}} - [Y_n]_{pv}[Y_n^\dagger]_{wr} \C_{\underset{vwst}{nu}} - \frac{1}{2}([Y^\dagger_u]_{wt}[Y_n]_{vr} \C_{\us{pvsw}{\ell nuq}} + [Y_u]_{sw}[Y^\dagger_n]_{pv} \C^{*}_{\us{rvtw}{\ell nuq}})\,,
\eeqa
\beqa
\dot \C_{\underset{prst}{\ell e}} &\supset& [Y_e]_{sr}\xi_{\us{pt}{e}} + [Y^\dagger_e]_{pt}\xi^*_{\us{rs}{e}} + [Y_n^\dagger Y_n]_{pr} \C_{\underset{st}{\phi e}} - [Y_n]_{pv}[Y_n^\dagger]_{wr} \C_{\underset{vwst}{ne}} \nonumber\\
&\quad&+\frac{1}{2}([Y_e]_{sw}[Y_n]_{vr} \C_{\us{pvwt}{\ell n\ell e}} + [Y^\dagger_e]_{wt}[Y^\dagger_n]_{pv} \C^{*}_{\us{rvws}{\ell n\ell e}})\,,
\eeqa
\beqa
\dot\C^{(1)}_{\us{prst}{qu}} \supset \frac{1}{N_c}[Y_u]_{sr}\xi_{\us{pt}{u}} + \frac{1}{N_c}[Y^\dagger_u]_{pt}\xi^*_{\us{rs}{u}}\,,
\eeqa
\beqa
\dot\C^{(8)}_{\us{prst}{qu}} \supset 2[Y_u]_{sr}\xi_{\us{pt}{u}} + 2[Y^\dagger_u]_{pt}\xi^*_{\us{rs}{u}}\,,
\eeqa
\beqa
\dot\C^{(1)}_{\us{prst}{qd}} \supset\frac{1}{N_c}[Y_d]_{sr}\xi_{\us{pt}{d}} + \frac{1}{N_c}[Y^\dagger_d]_{pt}\xi^*_{\us{rs}{d}}\,,
\eeqa
\beqa
\dot\C^{(8)}_{\us{prst}{qd}} \supset 2[Y_d]_{sr}\xi_{\us{pt}{d}} + 2[Y^\dagger_d]_{pt}\xi^*_{\us{rs}{d}}\,.
\eeqa

\begin{boldmath}
	\subsubsection{$(\bar LR)(\bar RL)$ and $(\bar LR)(\bar LR)$}
\end{boldmath}

\beqa
\dot \C_{\underset{prst}{\ell edq}} &\supset& -2[Y_d]_{st}\xi_{\us{pr}{e}} -2 [Y^\dagger_e]_{pr}\xi^*_{\us{ts}{d}} + 2 [Y^\dagger_n]_{pv}[Y_e^\dagger]_{wr} \C^{(1)*}_{\underset{wvst}{\ell nqd}} + 2 [Y^\dagger_n]_{pv}[Y_u]_{wt} \C_{\underset{vrsw}{nedu}}\,,
\eeqa
\beqa
\dot \C^{(1)}_{\underset{prst}{\ell equ}} &\supset& 2[Y^\dagger_u]_{st}\xi_{\us{pr}{e}} + 2 [Y^\dagger_e]_{pr}\xi_{\us{st}{u}} + 2 [Y^\dagger_n]_{pv}[Y_e^\dagger]_{wr} \C^{*}_{\us{wvts}{\ell nuq}} - 2 [Y^\dagger_n]_{pv}[Y_d^\dagger]_{sw} \C_{\us{vrwt}{nedu}}\,,
\eeqa
\beqa
\dot \C^{(3)}_{\underset{prst}{\ell equ}} &\supset& \frac{1}{2} [Y^\dagger_n]_{pv}[Y_d^\dagger]_{sw} \C_{\us{vrwt}{nedu}}\,,
\eeqa
\beqa
\dot\C^{(1)}_{\us{prst}{quqd}} \supset - 2[Y^\dagger_u]_{pr}\xi_{\us{st}{d}} - 2[Y^\dagger_d]_{st}\xi_{\us{pr}{u}}\,.
\eeqa
\section{Summary}
\label{sec:sum}
We presented the Yukawa terms of the one-loop anomalous dimension matrix for the dimension-six four-fermion operators of SMNEFT.
This complements the gauge terms calculated in Ref.~\cite{Datta:2020ocb}. Even if the right-handed neutrino Yukawa couplings are small, the induced 
mixings between SMNEFT operators can result in large RG running proportional to the other Yukawa couplings.
The Yukawa couplings of the right-handed neutrinos also cause the SMEFT and SMNEFT operators to mix.
We calculated the new ADM contributions for the 14 four-fermion SMEFT operators that are affected by this operator mixing.


\acknowledgments
 This work was supported by NSF Grant No. PHY1915142 (A.D.), Humboldt Foundation (J.K.), DOE Grant No. DE-FG02-95ER40896 and PITT PACC (H.L.), and DOE Grant No. de-sc0010504 (D.M.).

\begin{appendix}

\end{appendix}

\bibliography{ref_SMNEFT}
\end{document}